%% file: procl_preprint.tex
\documentstyle[procl]{article}

\input{psfig}

\def\Journal#1#2#3#4{{#1} {\bf #2}, #3 (#4)}

\def\NPB{{\em Nucl. Phys.} B}

\def\ZPC{{\em Z. Phys.} C}

\def\lsim{\mathrel{\raise.2ex\hbox{$<$}\hskip-.8em\lower.9ex\hbox{$\sim$}}}

\vspace*{-3cm}
\begin{document}
\hfill$\vtop{   \hbox{\normalsize DAPNIA / SPhN 96-20 }
                \hbox{\normalsize TTP96-29}
                   }$ 

\vspace*{5mm}
\title{PROSPECTS FOR MEASURING  $\Delta g$  FROM JETS 
AT HERA WITH POLARIZED PROTONS
}

\author{ J. FELTESSE, F. KUNNE }

\address{DAPNIA, CE Saclay, F91191 Gif/Yvette, France\\}

\author{ E. MIRKES }

\address{Inst. f. Theor. Teilchenphysik, 
         Universit\"at Karlsruhe, D-76128 Karlsruhe, Germany\\}




\maketitle\abstracts{
 The measurement of the polarized gluon distribution
 function $\Delta g(x)$  from photon gluon fusion processes in 
 electron proton deep inelastic scattering producing
 two jets has been investigated. The study is based on the MEPJET
 simulation program.
 The size of the expected  spin asymmetry
 and corresponding statistical uncertainties
 for a possible measurement
 with polarized beams of electrons and protons  at HERA 
 have been estimated.
 The results show that the asymmetry can reach a few percents.  
}  

\section{Introduction}

\input glue.tex


\end{document}

%% file: glue.tex
After confirmation of the surprising EMC result,
that quarks carry  only a little fraction of the 
nucleon spin,
this subject is  actively  being studied by several
fixed target experiments
at CERN, DESY and SLAC~.\cite{ro}
Sofar only the polarized structure functions
$g_1(x_g,Q^2)$ and $g_2(x_g,Q^2)$
have been measured.
 These do not allow to distinguish between
the role  of quarks and gluons distributions.
The measurement of the polarized gluon distribution 
$\Delta g(x_g,Q^2)$
has become the key experiment
in order to understand the 
QCD properties of the spin of the nucleon.
We study here the possible direct measurement of $\Delta g(x_g,Q^2)$
from jet events at the HERA collider,
in the scenario where both the electron and the proton
beam  are polarized.

\begin{figure}
\psfig{figure=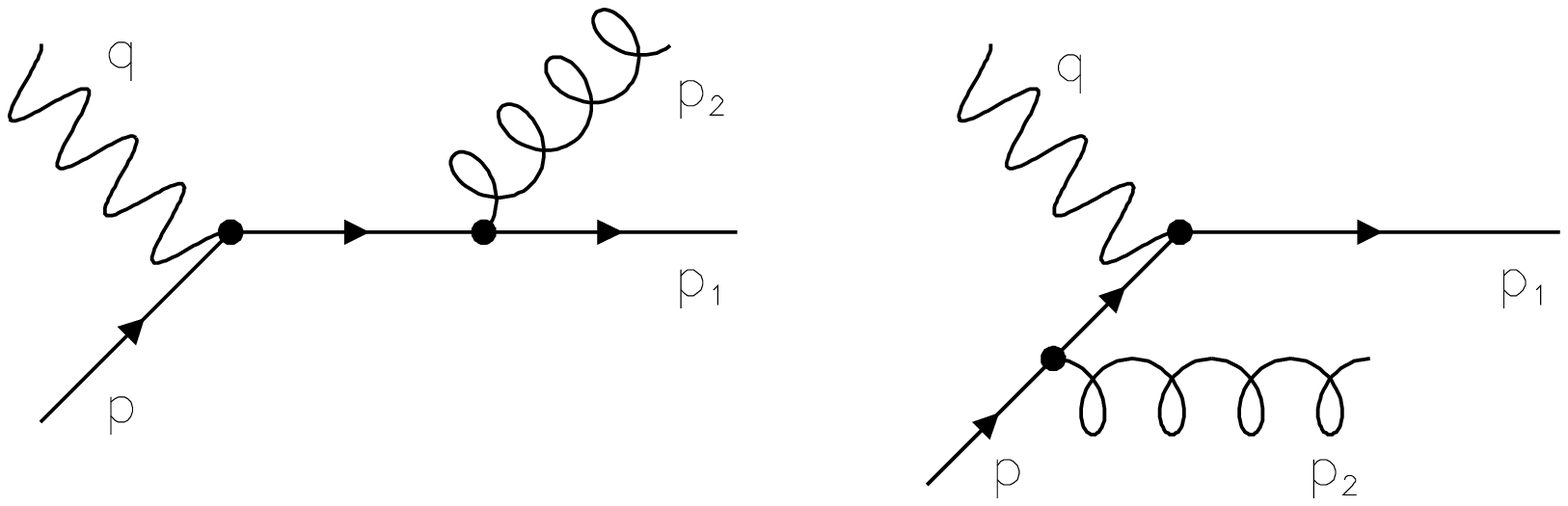,height=2.5in}
\vspace{-6.1cm}\hspace*{6cm}
\psfig{figure=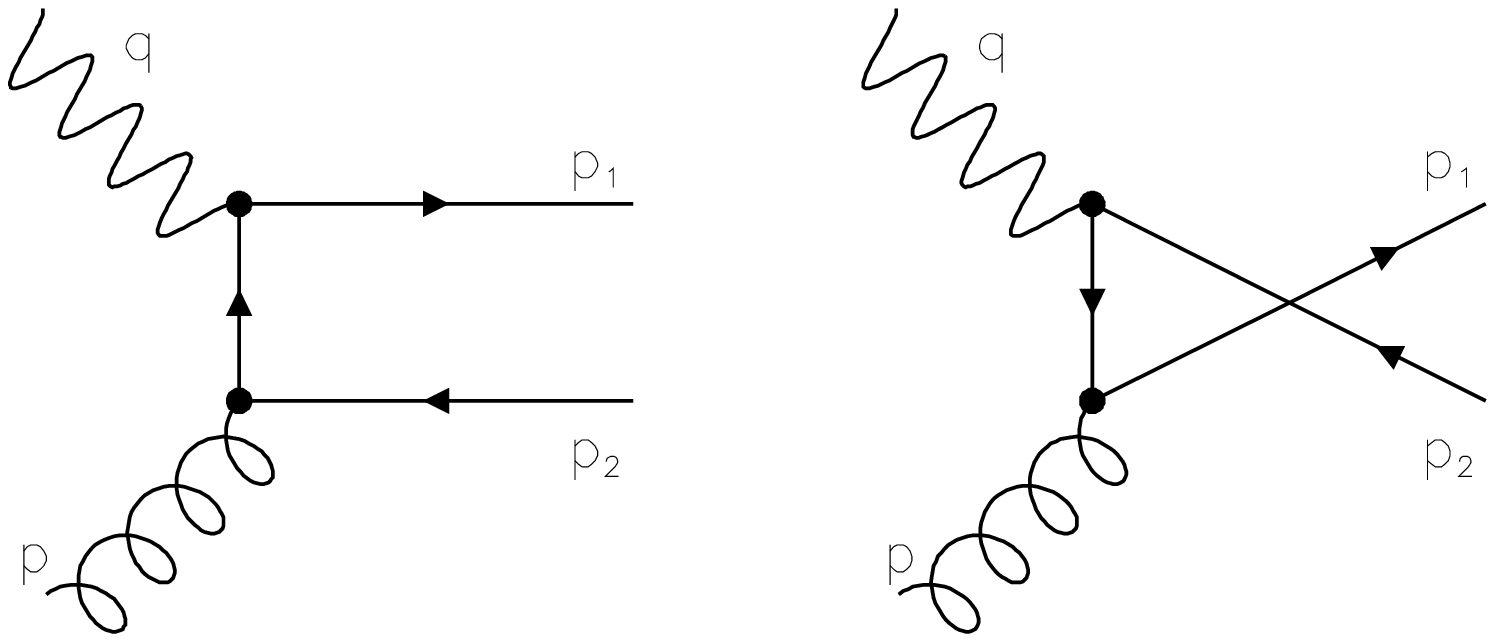,height=2.5in}
\caption{Feynman diagrams for the dijet cross section at LO.}
\label{fig:feyn}
\end{figure}

The gluon distribution  enters at leading order (LO)
in the two-jets production cross section \footnote{In the following 
the jet due to the beam remnant is not included in the number of jets.} 
in deep inelastic scattering (DIS) (see Fig.\ref{fig:feyn}),
and the unpolarized gluon distribution $g(x_g,Q^2)$
has indeed already been extracted from two-jets events
by the H1 collaboration at HERA.
With the  modest luminosity of  $0.24 ~pb^{-1}$
obtained in 1993, first data on
$x_g g(x_g)$ were extracted from  dijets events~\cite{h1}
at LO,  in  a wide  $x_g$ range $0.002 < x_g < 0.2$,
at  a mean $Q^2$ of $30 ~$ GeV$^2$.
These results  happen
to be in good agreement with the gluon distribution
extracted at LO from scaling violations of the structure function
$F_2$.

\section{Jet cross sections  in DIS  }
Deep inelastic electron proton scattering with several partons
in the final state,
\begin{equation}
e^-(l) + p(P) \rightarrow  e^-(l^\prime)+
\mbox{remnant}(p_r) +
\mbox{parton} \,\,1 (p_1) +
\ldots
+\mbox{parton}\,\, n (p_n)
\label{eq1}
\end{equation}
proceeds via the exchange of an
intermediate vector boson $V=\gamma, Z$. In the following,
$Z$-exchange will be neglected. We denote the 
momentum of the virtual photon, $\gamma^{\ast}$, by $q=l-l^\prime$, 
its absolute square by $Q^2$,
and use the standard scaling variables 
$x_{Bj}={Q^2}/({2P\cdot q})$ and $y={P\cdot q}/{P\cdot l}$.
The general structure of the {\it unpolarized}
 $n$-jet cross section in DIS is given by
\begin{equation}
d\sigma^{had}[n-\mbox{jet}] = \sum_a
\int dx_a \,\,f_a(x_a,\mu_F^2)\,\,\, d\hat{\sigma}^a(p=x_a P, 
\alpha_s(\mu_R^2), \mu_R^2, \mu_F^2)
\label{eq2}
\end{equation}
where the sum runs over incident partons $a=q,\bar{q},g$ which carry 
a fraction $x_a$ of the proton momentum.
$\hat{\sigma}^a$ denotes the partonic cross section from which collinear 
initial state singularities have been factorized out 
(in  next-to-leading order (NLO)) at a scale $\mu_F$ and 
implicitly included in the scale dependent parton densities 
$f_a(x_a,\mu_F^2)$. 
For {\it polarized} lepton hadron scattering, the hadronic ($n$-jet) 
cross section is obtained from Eq.~(\ref{eq2}) with the replacements
$(\sigma^{had}, f_a\,\, , \hat{\sigma}^a)\rightarrow
 (\Delta\sigma^{had}, \Delta f_a, \Delta\hat{\sigma}^a)$.
The polarized hadronic cross section is defined by
$\Delta\sigma^{had}\equiv\sigma^{had}_{\uparrow\uparrow}
                   -\sigma^{had}_{\uparrow\downarrow}$, where the
left arrow in the subscript denotes the polarization of the 
incoming lepton with respect to the direction of its momentum.
The right arrow stands for the polarization of the proton parallel 
or anti-parallel to the polarization of the incoming lepton.
The polarized parton distributions are defined by
$\Delta f_a(x_a,\mu_F^2)
\equiv f_{a \uparrow}(x_a,\mu_F^2)-f_{a \downarrow}(x_a,\mu_F^2)$.
Here, $f_{a \uparrow} (f_{a \downarrow})$ denotes a parton $a$ 
in the proton whose spin is aligned (anti-aligned) to the proton's spin.
$\Delta\hat{\sigma}^a$ is the corresponding polarized
partonic cross section.

In Born approximation, the subprocesses  
$\gamma^\ast +q \rightarrow q + g$, 
$\gamma^\ast +\bar q \rightarrow \bar q + g$,  
$\gamma^\ast+g \rightarrow q + \bar{q}$  
contribute to the two-jet cross section (Fig.\ref{fig:feyn}).
The boson gluon fusion subprocess
$\gamma^\ast+g \rightarrow q + \bar{q}$  dominates the
two jet cross section at low $x_{Bj}$ (see below) and allows for a  direct
measurement of the gluon density in the proton.
The full NLO corrections for two jet production in unpolarized
lepton hadron scattering  are now available~\cite{mi} 
and implemented in the 
$ep \rightarrow n$ jets event generator MEPJET,
which allows to analyze 
arbitrary jet definition schemes and 
general cuts in terms of parton 4-momenta.

First discussions about jet
production in polarized lepton-hadron scattering
can be found in Ref.~[4], where the jets were defined in
a modified ``JADE'' scheme. 
However, it was found \cite{mi} 
that  the theoretical uncertainties 
of the two-jet cross section for the ``JADE'' scheme can be very large
due to higher order effects.
These uncertainties are small for the cone scheme and the following 
results are therefore based on the cone algorithm,
which is defined in the laboratory frame.
In this algorithm  the distance 
$\Delta R=\sqrt{(\Delta\eta)^2+(\Delta\phi)^2}$ between two partons 
decides whether they should be recombined to a single jet. Here the variables 
are the pseudo-rapidity $\eta$ and the azimuthal angle $\phi$. We 
recombine partons with $\Delta R<1$.
Furthermore, a cut on the jet transverse momenta of $p_T(j)>5$~GeV in the lab 
frame and in the Breit frame is imposed.
We employ the one loop (two loop) formula for the strong coupling constant
in a LO (NLO) analyses
with 
a value for ${\Lambda_{\overline{MS}}^{(4)}}$
consistent with the value from the parton distribution functions.
In addition a minimal set of general  kinematical cuts
is imposed on the virtual photon and on the final state electron and jets.
If not stated otherwise, we require
5~GeV$^2<Q^2<2500$ GeV$^2$,
$0.3 < y < 1$, an energy cut of $E(e^\prime)>5$~GeV on the scattered 
electron, and a cut on the pseudo-rapidity $\eta=-\ln\tan(\theta/2)$
of the scattered lepton (jets) of $|\eta|<3.5$ ($|\eta|<2.8$).
These cuts are compatible with the existing
detectors H1 and ZEUS.

Let us briefly discuss the choice of the renormalization and factorization
scales $\mu_R$ and $\mu_F$ in Eq.~(\ref{eq2}).
Jet production is a multi-scale problem and it is not a priori clear at 
which scale $\alpha_s$ and the parton densities are probed.
However, it was argued \cite{rheinsberg,rom}, that the ``natural'' scale
for jet production in DIS
is set by the average $k_T^B$
of the jets in the Breit frame.
Here, $(k_T^{B}(j))^2$ is defined by 
$2\,E_j^2(1-\cos\theta_{jP})$, where the subscripts $j$ and $P$
denote the jet and proton, respectively (all quantities are defined
in the Breit frame).
It can be shown \cite{rheinsberg}
that $\sum_j \,k_T^B(j)$ smoothly interpolates 
between the correct limiting
scale choices, it approaches $Q$ in the parton 
model limit and it corresponds to 
the jet
transverse momentum (with respect to the $\gamma^*$-proton
direction) when the photon virtuality becomes negligible.
It therefore 
appears to be the ``natural'' scale for multi jet production in DIS
and we choose $ \mu_R^2 = \mu_F^2 = 1/4\;(\sum_j \,k_T^B(j))^2$
in Eq.~(\ref{eq2})
as our standard choice. 

Let us first discuss some results for unpolarized dijet cross sections.
If not stated otherwise,
the lepton and hadron beam energies are 27.5 and 820 GeV, respectively.
With the previous parameters and  GRV parton densities~\cite{grv}
one obtains a LO (NLO) two jet cross section $\sigma^{had}(2-\mbox{jet})$
of 1515 pb (1470 pb).
Thus the higher order corrections are small. This is  essentially due
to the relatively large cuts on the transverse momenta of the jets.
As mentioned before, the boson-gluon fusion subprocess 
dominates the cross section and contributes to 80\% to the LO cross section.

In order to investigate 
the feasibility of the parton density determination,
Fig.~2a  shows the Bjorken $x_{Bj}$ distribution of the
unpolarized two jet exclusive cross section.
The gluon initiated subprocess clearly dominates the Compton process
for small $x_{Bj}$ in the LO predictions. 
The effective $K$-factor close to unity
for the total exclusive dijet cross section is a consequence of
compensating effects in the low $x$ ($K>$ 1) and high $x$ ($K<$1) regime.

For the isolation of parton structure functions we are interested in the 
fractional momentum $x_a$ of incoming parton $a$ ($a=q,g$), however, and in
dijet production $x_{Bj}$ and  $x_a$ differ substantially. 
Denoting as ${s_{ij}}$
the invariant mass squared of the produced dijet system, and considering 
two-jet exclusive events only, the two are related by
%
%
$
x_a = x_{Bj} \,\left(1+\frac{{s_{ij}}}{Q^2}\right).
$
%
%
The $s_{ij}$ distribution of Fig.~2b exhibits rather large NLO
corrections as well. The invariant mass squared of the two jets is
larger at NLO than at LO (the mean value of $s_{ij}$ rising to 620~GeV$^2$ 
at NLO from 500~GeV$^2$ at LO).

The NLO corrections to the $x_{Bj}$ and $s_{ij}$ distributions
have a compensating effect on the $x_a$ distribution in Fig.~2c, which shows
similar shapes at LO and NLO. 
At LO a direct determination  of the gluon
density is possible from this distribution, after subtraction of the 
calculated Compton subprocess. This simple picture is modified in NLO,
however, and the effects of Altarelli-Parisi splitting and low $p_T$ partons
need to be taken into account more carefully to determine the structure
functions at a well defined factorization scale $\mu_F$ in NLO.

In the following we discuss some results for polarized dijet production.
Our standard set of polarized parton distributions is ``gluon, set A''
of Gehrmann and Stirling~\cite{gs},
for which $\int \Delta g = 1.8$ at $Q^2 = 4 $~GeV$^2$.
Using the same kinematical cuts as before, 
the LO polarized dijet cross section 
$\Delta{\sigma}(2-\mbox{jet})$ is -45 pb.
This negative result for the polarized dijet cross section is entirely due
to the boson-gluon fusion process, which is negative
for $x_{Bj}\lsim 0.025$. The  contribution to the total
polarized dijet cross section
is -53 pb. The contribution from the quark initiated subprocess 
is positive over the whole kinematical range and contributes with
8 pb to the resulting dijet cross section.
Note, however, that the shape of the
$x_g$ distribution in the polarized gluon density
is hardly (or even not at all)  constrained by currently 
available DIS data (in particular for small $x_g$).
Alternative parametrizations of the polarized gluon distributions in
the small $x_g$ region,
which are still consistent with all present data \cite{gs1},
can lead to 
polarized cross-sections $\Delta\sigma^{had}$,
which are at least a factor two larger
than those obtained with ``gluon, set A''.
It is clear from Fig.~2d, that the dijet events are particularly
sensitive to this lower $x_g$ range.
Note, that this fractional momentum distribution 
is again related to $x_{Bj}$
by
$
x_a = x_{Bj} \,\left(1+\frac{{s_{ij}}}{Q^2}\right).
$
The corresponding $x_{Bj}$ and $s_{ij}$ distributions 
are not shown here.


%
%
\begin{figure}[htb]
\psfig{figure=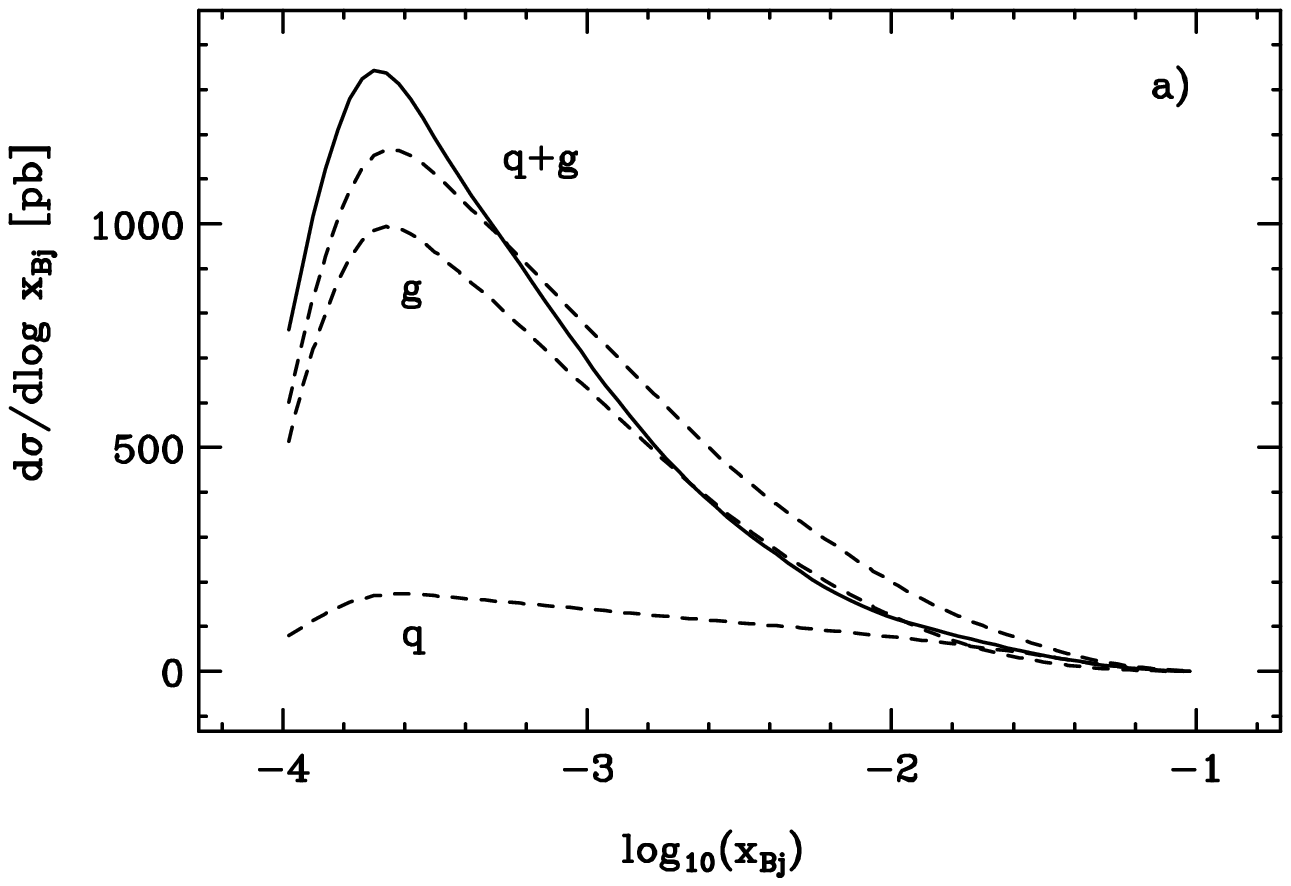,height=1.5in}
\vspace{-3.8cm}
\hspace*{4cm}
\psfig{figure=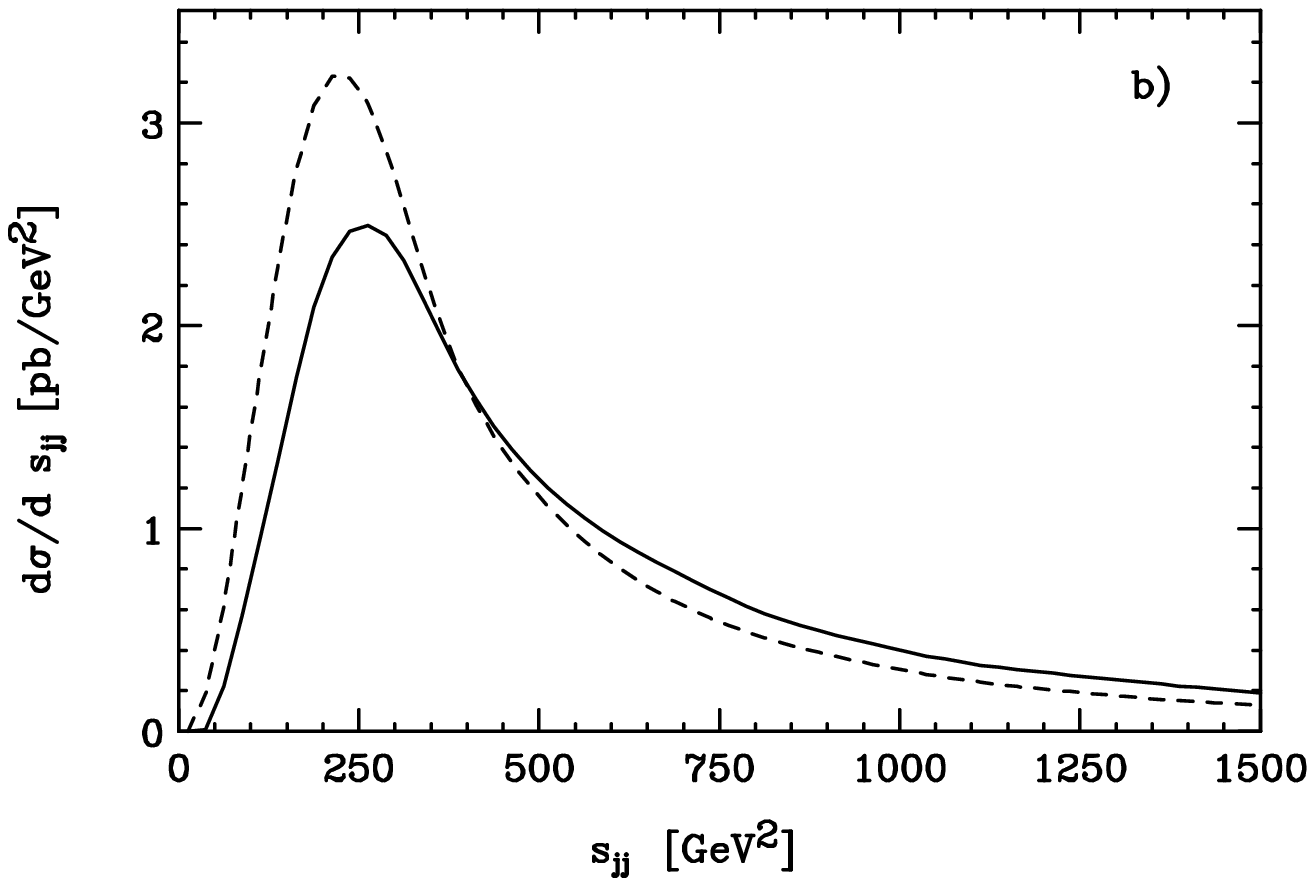,height=1.5in}
\vspace{2mm}
\psfig{figure=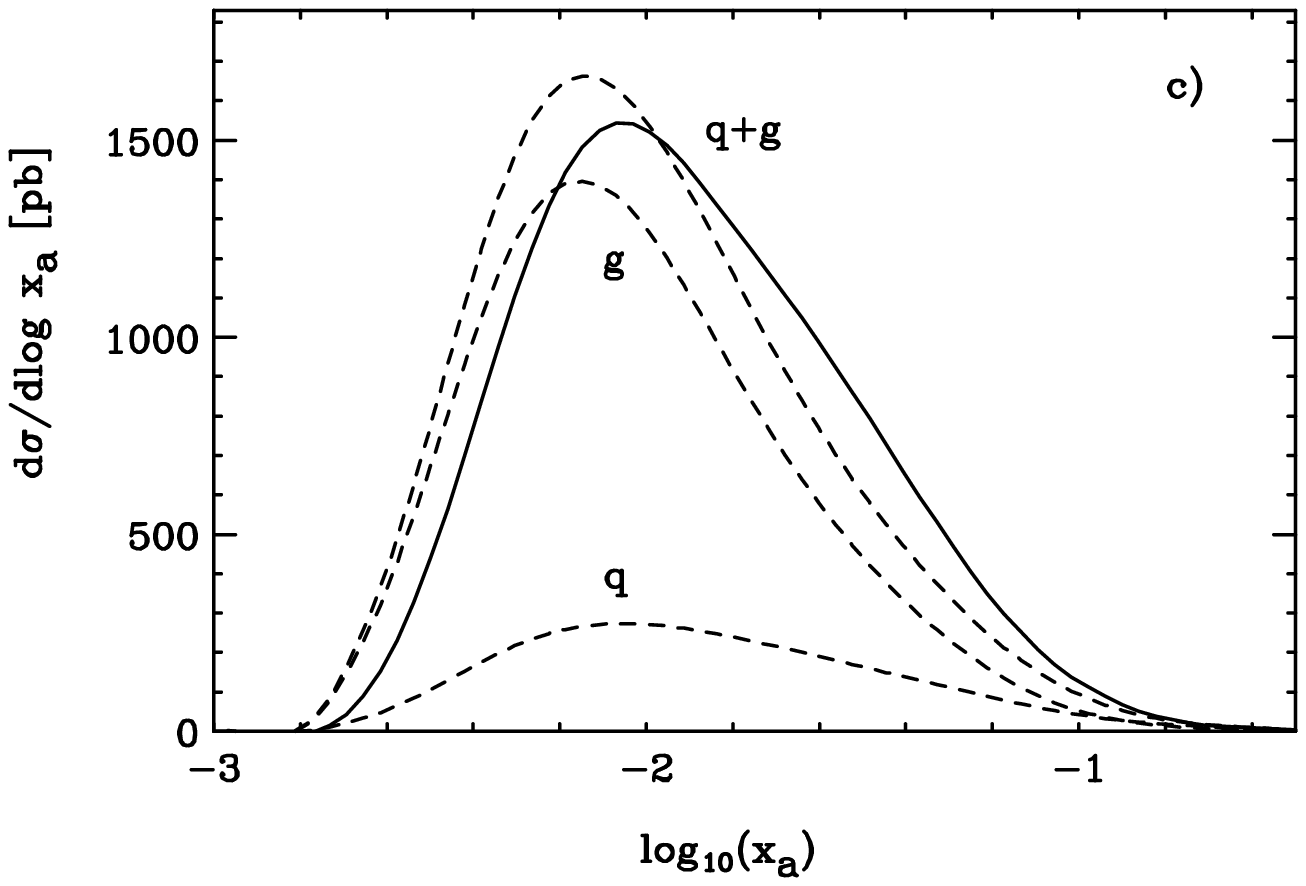,height=1.5in}\\[-4cm]
\hspace*{6.2cm}
\psfig{figure=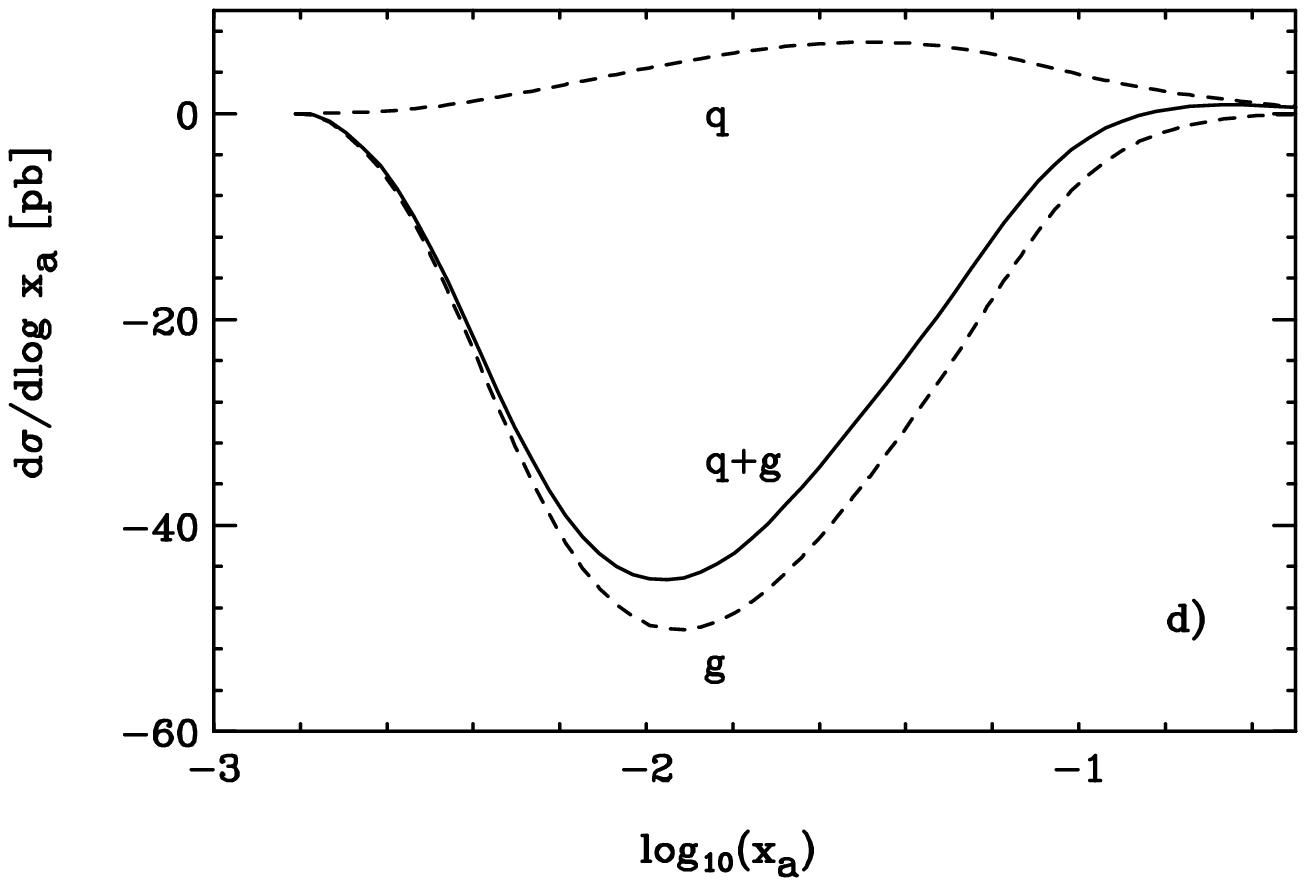,height=1.5in}
\caption{a) Dependence of the  {\it unpolarized}
two-jet cross section
on Bjorken $x_{Bj}$ for the
quark and gluon initiated subprocesses and for the 
sum. Both LO (dashed) and NLO (solid) results are shown;\,\,
b) Dijet invariant mass distribution in LO (dashed) and in NLO (solid)
for unpolarized dijet production;\,\,
c) Same as a) for the $x_a$ distribution, 
$x_a$ representing the 
momentum fraction of the incident parton at LO;\,\,
d)  Dependence of the LO {\it polarized}
two-jet cross section
on  $x_a$ for the
quark and gluon initiated subprocesses (dashed) and for the 
sum (solid).
}
\label{fig2}
\end{figure}

\section{Experimental asymmetries}
In order to study the feasibility and 
the sensitivity of the measurement of the
spin asymmetry at HERA, we have assumed
polarizations of 70\%  for both 
the electron and the proton  beams and
statistical errors were calculated for a total luminosity of
$200 ~pb^{-1}$.

\begin{figure}
\psfig{figure=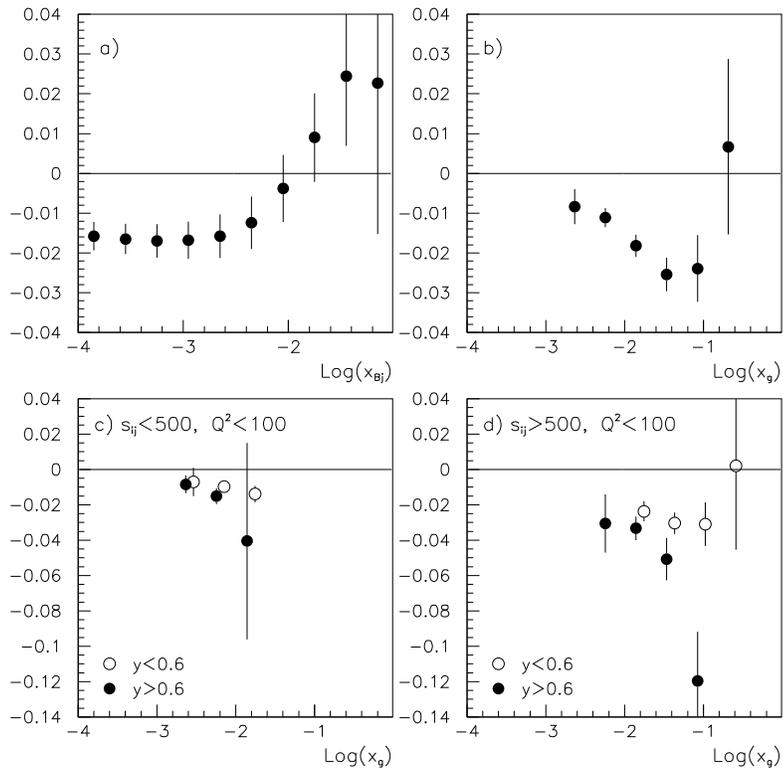,height=11cm}
\caption{Expected asymmetries as a function of $x_{Bj}$ (a)
and $x_g$ (b-d) for a luminosity of $200 ~pb^{-1}$
and beam polarizations $P_e = P_p = 70 \%$;
Fig.a) and b):nominal cuts;
Fig.c) and d): $Q^2 < 100 $~GeV$^2$;
In Fig.c) and d) data are separated in bins of $s_{ij}$ and $y$.
}
\label{fig:asy}
\end{figure}

The expected experimental asymmetry
$<A>
=\frac{\Delta\sigma^{had}(\mbox{2-jet})}{\sigma^{had}(\mbox{2-jet})}$
under these conditions is shown in
Fig.~\ref{fig:asy}a, as a function of $x_{Bj}$ and
in Fig.~\ref{fig:asy}b-d as a function of $x_g$.
Figures a) and b) correspond to the nominal kinematical cuts
defined previously, except for the $Q^2$ range which was extended
to lower values $2 < Q^2 < 2500$~GeV$^2$.
The cross section integrated over all variables is $2140 ~pb$,
where  82\% of the  contribution 
comes from gluon--initiated events. The
asymmetry averaged over all variables is $<A> = -0.015 ~\pm ~0.0015$.
It is negative at low $x_{Bj}$ and becomes positive
at $x_{Bj}>0.01$.

In figures c) and d), a further cut was made:
$Q^2 < 100$~GeV$^2$. 
This cut permits to reject the positive contributions to the asymmetry
coming from high $Q^2$ (equivalent to high $x_{Bj}$) events,
where the contribution of quark--initiated events is higher.
All the remaining events were separated in two
bins in $s_{ij}$ -the invariant mass of the dijet-
and two bins in $y$, as the asymmetry is
very sensitive to these two variables.
Fig. c)  corresponds to low invariant masses ($s_{ij} < 500$~GeV$^2$),
and Fig. d) to high ones ($s_{ij} > 500 ~$GeV$^2 $).
Open  points show low  $y$ values ( $y < 0.6$),
and closed points, high  $y$ values ( $y > 0.6$).
In the best case, the asymmetry reaches  values as high as
$12\%$ (Fig. d).

Reducing the beam energy to 410 GeV, instead of the nominal
820 GeV, does not improve the signal in average,
although the mean value of $y$ is higher.
The asymmetry signal increases only for a few points around $x_g>0.1$,
since a higher incident energy probes slightly higher values of $x_g$.

The  results show that if the assumed luminosity
and beam polarizations can be delivered at HERA,
the present detectors H1 and ZEUS will be in 
a comfortable position to measure a  spin asymmetry
of a few per cent in average,  with a
few per mil statistical precision.
In order to minimize the experimental systematic uncertainties,
it  is desirable  to have in the HERA ring
bunch trains of protons with alternate helicity.
On the theoretical side, NLO QCD corrections  are needed.
The NLO corrections reduce the renormalization $\mu_R$ and factorization scale
$\mu_F$  dependence 
(due to the initial state collinear factorization)
in the LO calculations and thus reliable predictions
in terms of a well defined strong coupling constant and
scale dependent parton distributions become possible.
At the moment, these corrections are only available for unpolarized
jet production \cite{mi,rom}.
One  expects for the asymmetry $<A>
=\frac{\Delta\sigma^{had}(\mbox{2-jet})}{\sigma^{had}(\mbox{2-jet})}$,
that the scale dependence in the individual cross sections
partly cancels in the ratio. 
In fact,  varying the renormalization and factorization scales
between
$ \mu_R^2 = \mu_F^2 = 1/16\;(\sum_j \,k_T^B(j))^2$
and
$ \mu_R^2 = \mu_F^2 = 4\;(\sum_j \,k_T^B(j))^2$
in the LO cross sections
introduces an uncertainty for the ratio $A$ of less than 2 \%,
whereas the uncertainty in the individual cross sections is much larger.

In  conclusion,  the dijets events  from polarized  electron
and polarized proton collisions at HERA can provide
a good measurement of the gluon polarization
distribution for $0.002 < x_g < 0.2$, the region
where $x_g\Delta g(x_g)$ is expected
to show a  maximum.
